\documentclass[preprint,amsmath,amssymb,showpacs]{revtex4}

\usepackage{graphicx} 
\usepackage{epsfig}
\usepackage{subfigure}

\begin{document}

\title{Exact corrections for finite-time drift and diffusion coefficients}
\author{C. Anteneodo}
\email{celia@fis.puc-rio.br}

\author{R. Riera}
\email{rrif@fis.puc-rio.br}

\affiliation{Department of Physics, PUC-Rio and  
National Institute of Science and Technology for Complex Systems,  
CP 38071, 22452-970, Rio de Janeiro, Brazil}

\begin{abstract}
Real data are constrained to finite sampling rates, which 
calls for a suitable mathematical description of the corrections 
to the finite-time estimations of the dynamic equations.
Often in the literature, 
lower order discrete time approximations of the modeling 
diffusion processes are considered. 
On the other hand, there is a lack of simple estimating procedures based 
on higher order approximations. 
For standard diffusion models, 
that include additive and multiplicative noise components,
we obtain the exact corrections to the empirical finite-time 
drift and diffusion coefficients, based on It\^o-Taylor expansions.
These results allow to reconstruct the real hidden coefficients from the empirical estimates.
We also derive higher-order finite-time expressions for the third and fourth 
conditional moments, that furnish extra theoretical checks for that class of diffusive models.
The theoretical predictions are compared with the numerical outcomes of some representative 
artificial time-series.
\end{abstract}


\maketitle

\section{Introduction}

Many fluctuating random phenomena, including turbulent diffusion, 
polymer dynamics or asset price evolution,
can be modeled by an univariate It\^o-stochastic differential equation (SDE) 
of the form bellow, characterizing a diffusive model:
\begin{equation} \label{ILE}
dX_t=D_1(X_t)dt+\sqrt{2D_2(X_t)}dW_t \,,
\end{equation}
where $W_t$ is a Wiener process, $D_1(X_t)$ is the coefficient of the slowly varying 
component (called drift coefficient) and $D_2(X_t)$ is the coefficient of the rapid 
 one (called diffusion coefficient).

For sufficiently smooth and bounded drift and diffusion coefficients, 
the associated probability density function (PDF)
 $P(x,t)\equiv P(X_t=x,t)$ 
is governed by the corresponding Fokker-Planck equation~\cite{risken}
\begin{equation} \label{FPE}
\partial_t P(x,t)= -\partial_x [D_1(x)P(x,t)]+ \partial_{xx}[D_2(x)P(x,t)] \,.
\end{equation}

Here, we are concerned with the empirical access to unknown drift and diffusion coefficients 
of stochastic processes. 
For an ideal time series $X_t$ generated by Eq.~(\ref{ILE}) and sampled 
with a sufficiently high resolution on a long time period, the original coefficients can be 
perfectly reconstructed.
For stationary processes, the coefficients $D_k(x)$, with $k=1,2$ can be directly estimated 
from the conditional moments~\cite{risken} as:
\begin{eqnarray} \label{ABs}
D_k(x)    &=& \lim_{\tau\to 0}\tilde{D}_k(x,\tau)\,,
\end{eqnarray}
where
\begin{eqnarray} \label{ABtilde}
\tilde{D}_k(x,\tau)  &=& \frac{1}{\tau\,k!}  \langle  [X_{t+\tau}-X_t]^k \rangle|_{X_t=x}    \,,
\end{eqnarray}
with $\langle \cdots\rangle$ denoting statistical average and 
$|_{X_t=x}$ meaning that at time $t$ the stochastic variable assumes the value $x$.

Conversely, for general Markovian stochastic processes, the time evolution of PDFs is 
governed by a generalization of Eq.~(\ref{FPE}), namely
\begin{equation} \label{KME}
\partial_t P(x,t)= \sum_{k\ge 0}(-\partial_x)^k [D_k(x)P(x,t)] \,.
\end{equation}
with coefficients $D_k(x)$ given by Eqs.~(\ref{ABs})-(\ref{ABtilde}), for any integer $k\ge 1$.
For diffusive processes, Eq.~(\ref{KME}) reduces to Eq.~(\ref{FPE}). 
Therefore, processes governed by the It\^o-Langevin Eq.~(\ref{ILE}) must furnish null 
coefficients $\tilde{D}_k$, for $k\ge 3$. 
Pawula theorem~\cite{risken} simplifies this task by 
stating that if $D_4$ is null, all other coefficients with $k\ge 3$ are null as well. 
The coefficient $D_4$ is then a key coefficient to be investigated, 
in order to establish the validity of the modeling of data series by Eq.~(\ref{ILE})-(\ref{FPE}).

However, due to the finite sampling rate of real data, numerical estimations of $\tilde{D}_k$ can 
not always be straightforwardly extrapolated to the limit $\tau\to 0$ in Eq.~(\ref{ABs}). 
In such cases, one accesses only the finite-$\tau$ estimation of the coefficients 
given by Eq.~(\ref{ABtilde}), which may significantly 
differ from the true coefficients ${D_k}$.
This is specially relevant when $\tau$ is large compared to the characteristic timescales 
of the process.

Some authors~\cite{friedrich4} have introduced finite sampling rate corrections 
to the coefficients ${D_1}(x,\tau)$ and ${D_2}(x,\tau)$, 
by deriving  expansions for the conditional moments up to some specified 
low order of $\tau$, directly from the Fokker-Planck equations. 
Applications of this approach have already been implemented for those coefficients 
up to second order~\cite{Julia}. 
The error in the finite-$\tau$ estimated coefficients $\tilde{D}_k$ can also be derived 
from the stochastic 
It\^o-Taylor expansion~\cite{book} of the integrated form of Eq.~(\ref{ILE}).  
Within this line, the first order expansion of drift and diffusion coefficients 
was recently presented in Ref.~\cite{sura}. 
However, low order corrections may be inappropriate 
when the convergence of the limit in Eq.~(\ref{ABs}) is slow~\cite{friedrich4,comment}.  
Moreover, there is no {\em a priori} knowledge of whether the 
sampling rate is fine enough to justify the use of the lowest order approximation.

In the present work, we investigate those issues for diffusion models defined by Eq.~(\ref{ILE})  
with linear drift coefficient, namely, ${D_1}(x)=-a_1x$, 
representing an harmonic restoring mechanism, and quadratic state-dependent diffusion coefficient, 
namely, ${D_2}(x)=b_0+b_2x^2$.
This class encompasses some of the most common models of the theoretical literature.
In fact, this equation is frequently found in a diversity of processes, from turbulence 
to finance~\cite{turbulence,turbfinance}.
Moreover, the obtained results are also valid for another class of SDEs 
with additive-multiplicative noises~\cite{multiplicative1,multiplicative2}, given by 
\begin{equation} \label{ILE2}
dX_t=-a_1X_tdt+\sqrt{2b_0}dW_t +\sqrt{2b_2}X_tdW^\prime_t \,,
\end{equation}
where $W_t,\,W^\prime_t$ are uncorrelated Wiener processes.

For discretely sampled data at intervals $\tau$, we will
derive, from the stochastic It\^o-Taylor expansion, 
finite-$\tau$ expressions for the parameters $\{{a}_1,{b}_0,{b}_2\}$, 
up to infinite order. 
These exact expressions will  allow us to reconstruct the true drift and 
diffusion coefficients from 
their  empirical finite-time estimates. 
As a corollary, one can determine up to which value of $\tau$ a given order of 
truncation is reliable (within a fixed tolerance), or reciprocally, 
which is the sufficient order for a given $\tau$.

Furthermore, as empirical estimates suffer from finite-$\tau$ effects, one always gets 
non-null $D_4$. Therefore, the evaluation of the corrections for this coefficient is 
crucial for a suitable probe of the diffusive modeling.
In this work, we also derive finite-$\tau$ expressions for coefficients $D_3$ and $D_4$, 
which furnish extra theoretical tests of consistency for the diffusive models considered.

Our theoretical findings are corroborated by the outcomes of exemplary artificial 
time-series generated by Eq.~(\ref{ILE}).

\section{Exact corrections  for drift and diffusion coefficients }

Let us consider the It\^o formula~\cite{book}, for a given function $F$ of the stochastic 
variable $X_t$
\begin{eqnarray} \nonumber
dF&=&(\partial_t F+D_1\partial_X F +D_2\partial_{XX} F )dt
+ \sqrt{2D_2}\partial_X F dW \\\label{L0L1}
&\equiv& L^0Fdt+L^1FdW,
\end{eqnarray} 
and its integrated form
\begin{equation} \label{intFform}
F(X_{t+\tau})=F(X_t) + \int_t^{t+\tau}L^0 F(X_s)ds+\int_t^{t+\tau}L^1 F(X_s)dW_s\,.
\end{equation} 

Let $\tau$ be the sampling interval of state space observations.
By applying It\^o formula (\ref{intFform}) to the functions $D_1(X_s)$ and $\sqrt{2D_2(X_s)}$ 
in the integral form of Eq.~(\ref{ILE}): 
\begin{equation} \label{intXform}
X_{t+\tau}=X_t +\int_t^{t+\tau} D_1(X_s)ds+\int_t^{t+\tau} \sqrt{2D_2(X_s)}dW_s,
\end{equation} 
one finds
\begin{eqnarray}\nonumber
X_{t+\tau}&=& X_t +\int_t^{t+\tau}\biggl( 
D_1(X_t)+\int_t^{s}L^0 D_1(X_s')ds' +  
 \int_t^{s} L^1 D_1(X_{s'})dW_{s'}\biggr)ds \\ 
 &+& \int_t^{t+\tau}\biggl( \sqrt{2D_2(X_t)}+  
 \int_t^{s}L^0 \sqrt{2D_2(X_s')}ds' +\int_t^{s} L^1 
\sqrt{2D_2(X_{s'})}dW_{s'}\biggr)dW_s\,.\label{expansion}
\end{eqnarray} 
After iterated applications of It\^o formula, 
one gets an expression for the increment of the 
stochastic variable in terms of multiple stochastic integrals~\cite{book}:
\begin{equation} \label{MSI}
X_{t+\tau}-X_t=\sum_{\alpha_k} c_{\alpha_k}(D_1,D_2)\,I_{\alpha_k},
\end{equation}
where $\alpha_k=(j_1,j_2,\ldots,j_k)$, with $j_i=0,1$ for all $i$, 
$c_{\alpha_k}(D_1,D_2)=L^{j_1}L^{j_2}\ldots L^{j_{k-1}}L^{j_k}$  and 
$I_{\alpha_k}$ are multiple stochastic integrals of the form 
$I_{\alpha_k}=\int_t^{t+\tau} \int_t^{t+t_{k}} \int_t^{t+t_{k-1}} 
 \ldots \int_t^{t+t_2} dt_1^{j_1}\ldots dt_{k-1}^{j_{k-1}} dt_{k}^{j_k}$, 
 with $dt_i^0\equiv dt_i$ and $dt_i^1\equiv dW_i$.

By inserting Eq.~(\ref{MSI}) into Eq.~(\ref{ABtilde}) and performing the averaging,  
for $k=1,2$, we achieve 
analytical expressions for the finite-$\tau$ drift and diffusion 
coefficients, up to arbitrary order in powers of $\tau$. 
The resulting expressions preserve the linear and quadratic 
$x$-dependence, respectively and can be written as:
\begin{eqnarray} \nonumber
\tilde{D}_1(x,\tau) &=& -\tilde{a}_1(\tau)x \\ \label{D1D2til} 
\tilde{D}_2(x,\tau) &=&  \tilde{b}_0(\tau)+\tilde{b}_2(\tau)x^2 \,.
\end{eqnarray}

Hence, we are led to the theoretical relation between the finite-$\tau$ coefficients 
$\{\tilde{a}_1,\tilde{b}_0,\tilde{b}_2\}$ and the
true ones $\{{a}_1,{b}_0,{b}_2\}$, namely, 
\begin{eqnarray}\label{a1n}
\tilde{a}_1(\tau)&=&  a_1\sum_{j\ge 0} \frac{[-a_1]^j}{(j+1)!}\tau^j  \,,\\ \label{b0n}
\tilde{b}_0(\tau) &=&  b_0\sum_{j\ge 0} \frac{[(-2(a_1-b_2)]^j}{(j+1)!}\tau^j \,,\\ \label{b2n}
\tilde{b}_2(\tau) &=& \sum_{j\ge0} \frac{ \frac{1}{2}[-2(a_1-b_2)]^{j+1} -[-a_1]^{j+1} }{(j+1)!} 
\tau^j \,.
\end{eqnarray}
Details of the derivation of Eqs.~(\ref{a1n})-(\ref{b2n})
can be found in the Appendix. 
 
By restricting the expansions~(\ref{a1n})-(\ref{b2n}) to some common finite power $n$, 
one gets the respective $n$th-order approximation.
This result extends previous findings of first~\cite{sura} and second~\cite{Julia} 
order  terms.

Notice that Eq.~(\ref{a1n}) is  uncoupled, meaning that the estimated  
harmonic stiffness $\tilde{a}_1$ is not affected by the exact noise components.
Moreover, from  Eq.~(\ref{b2n}), the estimated multiplicative noise parameter 
$\tilde{b}_2$ does not depend on the exact additive noise component. 

Summing the series in Eqs.~(\ref{a1n})-(\ref{b2n}) up to infinite order, 
and defining $Z\equiv \exp(-a_1\tau)$ and $W\equiv \exp(-2 b_2 \tau)$, 
we find the exact finite-$\tau$ expressions:
\begin{eqnarray} \nonumber  
\tilde{a}_1 &=&   \frac{1-Z}{\tau} \,,\\ \nonumber  
\tilde{b}_0 &=&  \frac{b_0}{a_1 - b_2} \,\frac{1-Z^2 W}{2\tau} \,,\\ \label{abtils} 
\tilde{b}_2 &=&  \frac{1-Z}{\tau} \,-\, \frac{1-Z^2 W}{2\tau} \,.
\end{eqnarray}
Notice that  $\lim_{\tau\to 0}  \{\tilde{a}_1,\tilde{b}_0,\tilde{b}_2\} =\{{a}_1,{b}_0,{b}_2\}$ 
holds.

From Eqs.~(\ref{abtils}), 
we obtain an invariant relation among the estimated and exact parameters, namely, 
\begin{equation} \label{constraint}
\frac{\tilde{a}_1 - \tilde{b}_2}{\tilde{b}_0} = \frac{a_1 - b_2}{b_0}\,.
\end{equation}

The meaning of this invariance can be drawn,  for instance, from the stationary 
PDF ${P^*}(x)$ associated to the corresponding Fokker-Planck equation 
given by Eq.~(\ref{FPE}). With the present choice of drift and diffusion coefficients, 
 for $a_1 , b_0 >0$, $b_2 \ge 0$, one has:
\begin{equation} \label{pdf}
{P^*}(x)\,=\,P_o/[1+\frac{b_2}{b_0}x^2]^{\frac{a_1}{2b_2}+1} \,,
\end{equation}
with $P_o$ a normalization constant.
This solution  is of the $q$-Gaussian form~\cite{multiplicative1}, for which, 
if ${a}_1-{b}_2>0$, the variance is finite with value $\sigma^2=b_0/(a_1-b_2)$.
Hence, Eq.~(\ref{constraint}) represents the uphold of the data variance under 
changes of sampling intervals. 
For ${b_2}=0$, one recovers the Gaussian stationary solution and its variance relation. 
Notice also that, from Eqs.~(\ref{a1n})-(\ref{b2n}), Eq.~(\ref{constraint})  
still holds if one considers partial corrections of the parameters up to any 
common order  $n$ of truncation of the sums.

Let us remark that the results presented in Eqs.~(\ref{D1D2til})-(\ref{abtils}) 
are valid even when the variance is infinite. 
However,  we will deal only with finite variance cases and 
consider normalized data (with unitary variance), 
which only implies a rescaling of $b_0\to b_0/\sigma^2 $. 
Then,  
\begin{equation} \label{constraint2}
{a}_1={b}_0+{b}_2 \,. 
\end{equation}
 
Taking into account the constraint~(\ref{constraint2}), from Eqs.~(\ref{abtils}),
the exact finite-$\tau$ expressions are:
\begin{eqnarray} \label{a1tilnor}
\tilde{a}_1 &=&  \frac{1-\exp(-a_1\tau)}{\tau} \\ \label{b0tilnor}
\tilde{b}_0 &=&  \frac{1-\exp(-2b_0\tau)}{2\tau}  
\,.
\end{eqnarray}

Eqs.~(\ref{a1tilnor}) and (\ref{b0tilnor}) can be readily inverted to extract the 
true parameters from their finite-$\tau$ estimates:
\begin{eqnarray} \label{a1}
{a}_1 &=&  \frac{\ln( 1 - \tilde{a}_1 \tau))}{-\tau}  \,,\\ \label{b0}
{b}_0 &=&  \frac{\ln( 1 - 2\tilde{b}_0 \tau))}{-2\tau}   \,.
\end{eqnarray}
Notice that $\tilde{a}_1\tau$ (and also $2\tilde{b}_0\tau$) 
can not be greater than unit.

In what follows, we fix the timescale $\tau=1$. 
A different choice would simply lead to a rescaling of the parameters 
$(a_1,b_0,b_2)\to (\tau a_1,\tau b_0,\tau b_2)$\,.

\begin{figure}[h!]
\centering
\includegraphics*[bb=120 265 520 830, width=0.6\textwidth]{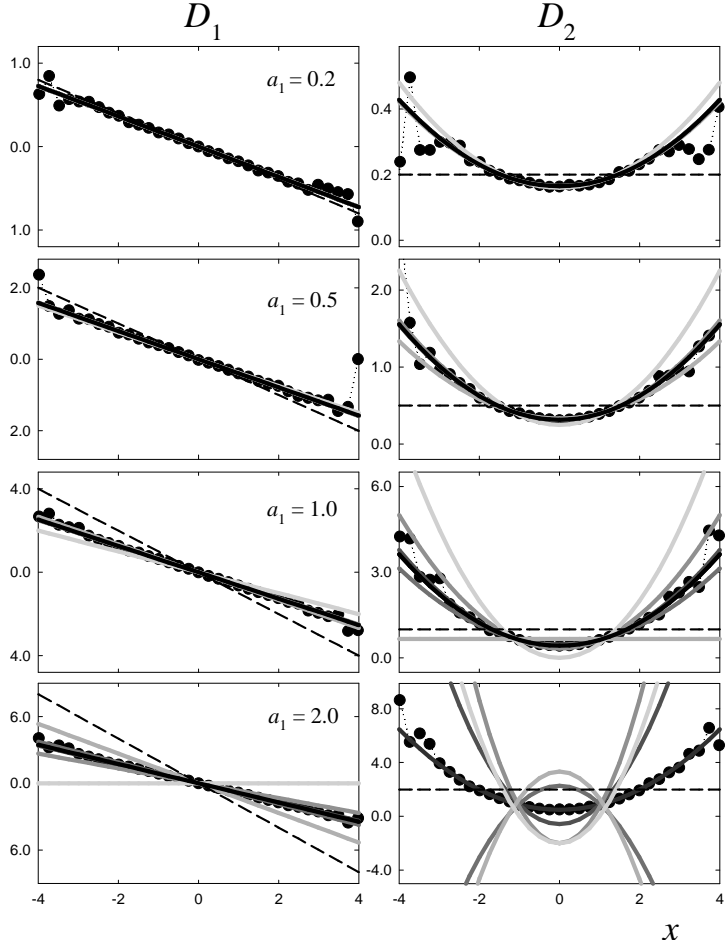}    
\caption{Drift and diffusion coefficients for the O-U process. 
Symbols correspond  to the numerical computation for artificial series ($10^5$ data), 
synthetized with the values of $a_1=b_0$ ($b_2=0$, in accord with constraint 
(\ref{constraint2})) indicated on each panel.
Lines represent the coefficients given by Eqs.~(\ref{D1D2til}), using 
the theoretical $\tau$-expansions~(\ref{a1n})-(\ref{b2n}), 
at different orders of truncation. The darker the color, the higher the order, 
from first up to fifth order. 
The infinite order (exact expression) is represented in thick black lines. 
The zeroth order,  corresponding to the true values, is plotted in dashed lines.
}
\label{fig:b2nullD1D2}
\end{figure} 

Now we investigate the importance of finite-$\tau$ effects for 
discretely sampled realizations of representative known diffusive processes. 
To this end, we generated artificial time-series through numerical integration of Eq.~(\ref{ILE}),
by means of an Euler algorithm with timestep $dt=10^{-3}$, 
recording the data at each $1/dt$ timesteps, in accord with our choice $\tau=1$. 
Our theoretical results  for ${D_1}$ and ${D_2}$ will be compared 
to the ones numerically computed from the time-series, through  Eq.~(\ref{ABtilde}).  

The particular case $b_2=0$, corresponding to the Orstein-Uhlenbeck (O-U) process 
and the general case with  multiplicative component $b_2>0$ will be 
investigated separately. 
Fig.~\ref{fig:b2nullD1D2} shows the results 
for the artificial series with known values of the parameters $a_1 = b_0$ ($b_2=0$), 
together with our theoretical predictions.  
The exact theoretical expressions reproduce the numerical (finite-time) outcomes.
Comparing  the panels in Fig.~\ref{fig:b2nullD1D2}, it is clear that,   
the larger $a_1$, the slower the convergence to the observed coefficients. 
The results for $a_1 > 1$ also illustrate the entanglement one may find 
in large-$\tau$ measurements, specifically, an oscillatory convergence of 
$\tilde {a}_1$ and an alternating signal of $\tilde {b}_2$.

\begin{figure}[hb!]
\centering
\includegraphics*[bb=110 35 520 780, width=0.6\textwidth]{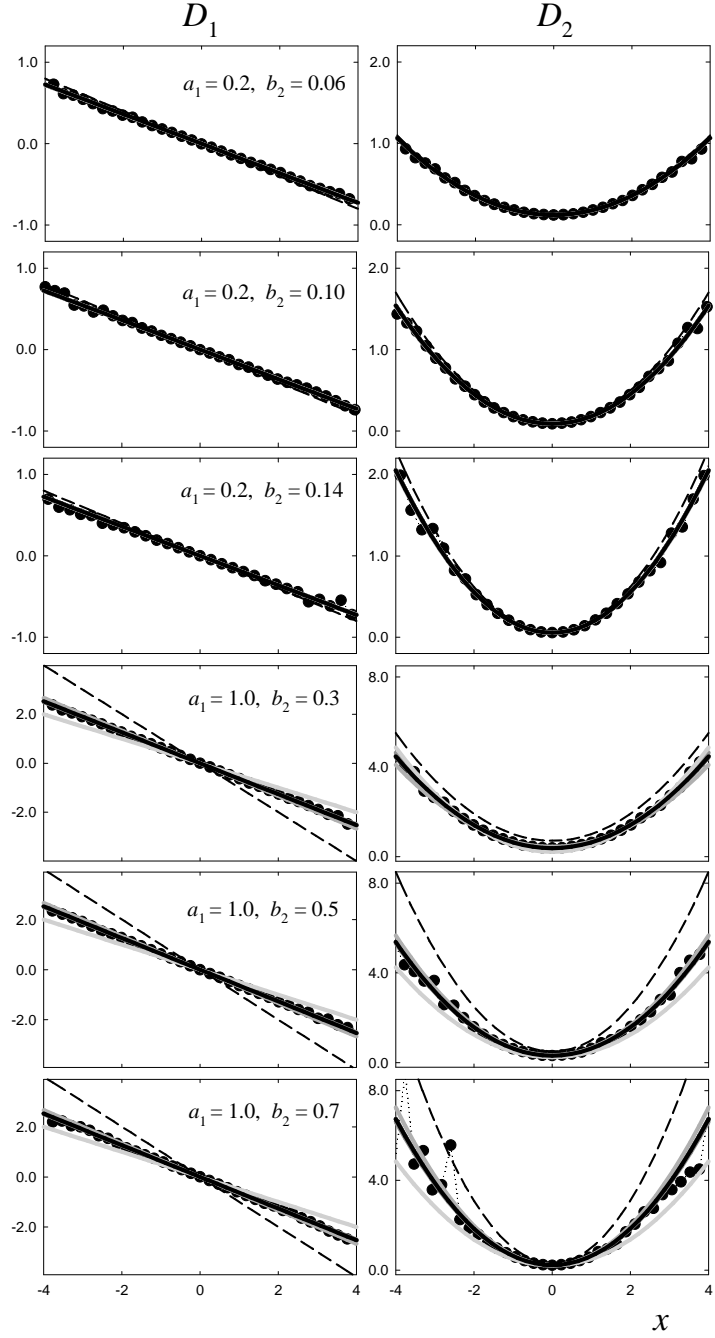}    
\caption{Drift and diffusion coefficients for the general process with multiplicative noise. 
Symbols correspond  to the numerical computation for artificial series ($10^6$ data), 
synthetized with the values of $a_1\; (=b_0+b_2$, in accord with constraint 
(\ref{constraint2}))  and $b_2$ indicated on each panel. 
Lines are as in Fig.~\ref{fig:b2nullD1D2}.
}
\label{fig:b2D1D2}
\end{figure}

In Fig.~\ref{fig:b2D1D2}, we plot the numerical computations for 
artificial time-series together with analytical predictions for $b_2>0$.
Again, the theoretical approximations present slower convergence as $a_1$ 
increases while the exact theoretical expressions agree with finite-time 
estimates directly obtained from the time-series.
Moreover, the actual value of $b_2$ sets the convergence rate of $\tilde {b}_2$.

\begin{figure}[h!]
\centering  
\includegraphics*[bb=134 210 507 665, width=0.5\textwidth]{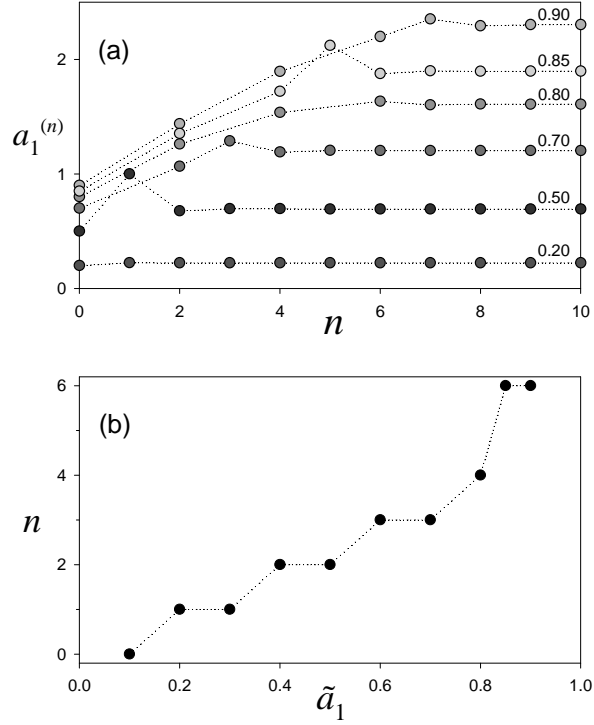}    
\caption{Dependence of ${a}^{(n)}_1$ on the order $n$ of the approximation given by 
Eq.~(\ref{a1n}), for different values of $\tilde{a}_1$ (panel (a)). Dotted lines correspond to the 
respective true values ($a_1$) and missing points denote 
the absence of real solutions. Panel (b) exhibits the 
order at which the limiting value is attained (within 5\%) as a function of $\tilde{a}_1$. 
}
\label{fig:orders}
\end{figure}

All these results raise a question about the domain of  validity of lower order approximations 
presented before in the literature. Let us investigate this issue quantitatively.  
Given $\tilde{a}_1$, obtained from numerical (finite-time) evaluation, the exact value of $a_1$ can 
be recovered from Eq.~(\ref{a1}). 
Approximate values $a_1^{(n)}$ can be obtained by inversion of Eq.~(\ref{a1n}) 
truncated at order $n$. 
Figure~(\ref{fig:orders}) illustrates $a_1^{(n)}$ as a function of 
$n$, for different values of $\tilde{a}_1$.
Clearly, convergence to the true value $a_1$  is attained (within  a given tolerance), 
at different orders that depend  on 
the value of $\tilde{a}_1$. 
For instance, for $\tilde{a}_1>0.5$, an order larger than two  is required.   
Convergence is faster for smaller $\tilde{a}_1$, that is, as soon as 1/$\tilde{a}_1$ becomes 
large compared to the timescale $\tau=1$. 
For $\tilde{b}_0$ we obtained a very similar convergence scheme (not shown).
In Refs.~\cite{sura,Julia}, the 
fitness of low order expressions for O-U processes 
results from the particular employment of $\tilde{a}_1\tau<0.5$. 
However, this may not be the case when dealing with generic empirical data. 
  
\section{Higher-order coefficients}

Inserting the It\^o-Taylor expansion Eq.~(\ref{MSI}) into Eq.~(\ref{ABtilde}) and 
performing the average for $k=3,4$, we also computed the 
finite-$\tau$ expansion for $\tilde{D}_3(x,\tau)$ and $\tilde{D}_4(x,\tau)$. 
The resulting expressions are invariant functions of $x$, namely:
\begin{eqnarray} \label{D3t}
\tilde{D}_3(x,\tau) &=& -\tilde{c}_1(\tau)+ \tilde{c}_3(\tau)x^3 \\ \label{D4t}
\tilde{D}_4(x,\tau) &=&  \tilde{d}_0(\tau)+\tilde{d}_2(\tau)x^2 + \tilde{d}_4(\tau)x^4\,.
\end{eqnarray}

For the particular case $b_2=0$, we were able to derive the infinite order expansion for 
the $\tau$-parameters:

\begin{eqnarray} \nonumber  
\tilde{c}_1(\tau) &=&  b_0\sum_{j\ge 0} \frac{3^{j+1}-2^{j+1}-1}{2}\frac{[-a_1]^j}{(j+1)!}\tau^j  
\,, \nonumber \\  
\tilde{c}_3(\tau)&=&  \sum_{j\ge 0} \frac{3^{j}-2^{j+1}+1}{2}\frac{[-a_1]^{j+1}}{(j+1)!}\tau^j \,,
\label{ctils}  \\  
\tilde{d}_0(\tau) &=& b_0^2\sum_{j\ge0} \frac{4^{j}-2^{j}}{2}\frac{ [-a_1]^{j-1}}{(j+1)!}  \tau^j \,,
\nonumber \\  
\tilde{d}_2(\tau) &=&    b_0\sum_{j\ge0} \frac{2\times4^{j}-3^{j+1}+1}{2}\frac{[-a_1]^{j} }{(j+1)!}  \tau^j\,,
\nonumber \\  
\label{dtils} 
\tilde{d}_4(\tau) &=&       \sum_{j\ge0} \frac{ 4^{j}-3^{j+1}-1}{6}\frac{[-a_1]^{j+1}}{(j+1)!} \tau^j \,.
\end{eqnarray}
Notice that, $\lim_{\tau\to 0}  \{\tilde{c}_1,\tilde{c}_3,\tilde{d}_0,\tilde{d}_2,\tilde{d}_4\} =0$\,, 
as expected, and that  the relevant parameter for the rate of series convergence is $a_1$.  
Summing the series (\ref{ctils})-(\ref{dtils}) up to infinite order, and 
recalling that $Z\equiv \exp(-a_1\tau)$, one obtains  
\begin{eqnarray} \nonumber  
\tilde{c}_1(\tau) &=& -\frac{b_0}{a_1} \frac{(1-Z)^2(1+Z) }{2\tau}   \,,\\ \label{ctilinf}  
\tilde{c}_3(\tau) &=&  -  \frac{(1-Z)^3 }{6\tau}\,,\\ \nonumber
\tilde{d}_0(\tau) &=& \frac{b_0^2}{a_1^2} \frac{(1-Z)^2(1+Z)^2}{8\tau} \,,\\ \nonumber  
\tilde{d}_2(\tau) &=& \frac{b_0}{a_1} \frac{(1-Z)^3(1+Z)}{4\tau}  \,,\\ \label{dtilinf}  
\tilde{d}_4(\tau) &=& \frac{(1-Z)^4}{24\tau} \,.
\end{eqnarray}

\begin{figure}[h!]
\centering
\includegraphics*[bb=110 255 520 780, width=0.6\textwidth]{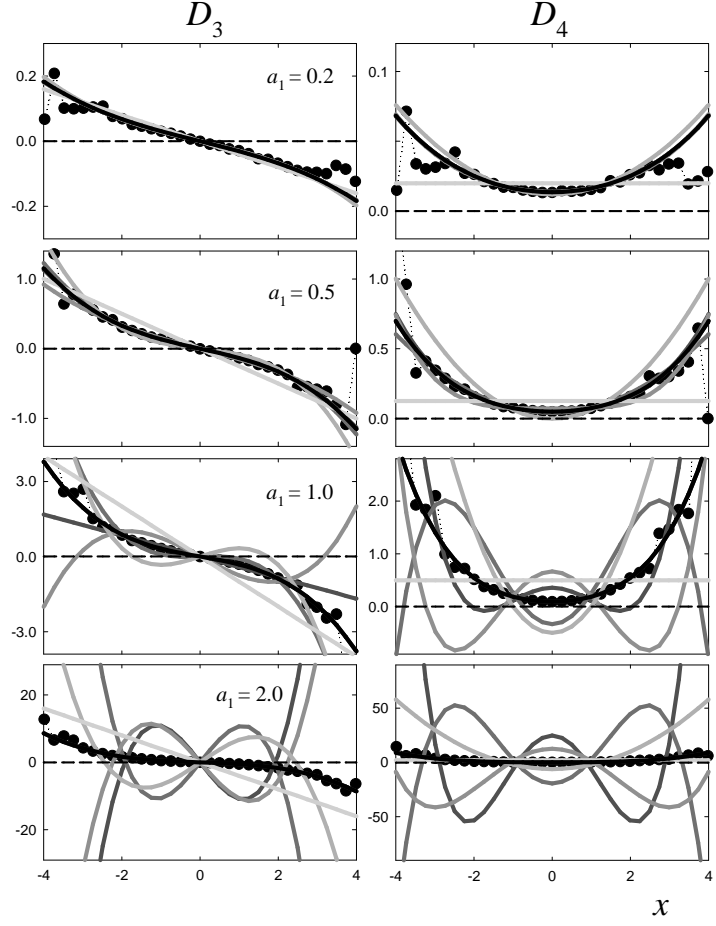}     
\caption{Third and fourth order coefficients for the O-U process. 
Symbols correspond  to the numerical computation for the same artificial series of 
Fig.~\ref{fig:b2nullD1D2}. 
Lines represent the coefficients given by Eqs.~(\ref{D3t})-(\ref{D4t}), using 
the theoretical $\tau$-expansions~(\ref{ctils})-(\ref{dtils}), 
at different orders of truncation. Colors as in Fig.~\ref{fig:b2nullD1D2}. 
}
\label{fig:b2nullD3D4}
\end{figure}

Fig.~\ref{fig:b2nullD3D4} shows the numerical computation of $\tilde{D}_3(x,\tau)$ and $\tilde{D}_4(x,\tau)$ 
for the same artificial series as in Fig.~\ref{fig:b2nullD1D2}. 
For comparison, the theoretical estimates at different orders of truncation 
of the series in Eqs.~(\ref{ctils})-(\ref{dtils}) are shown.
Notice that, although  ${D}_3,{D}_4=0$ for the diffusive processes considered here, 
their finite-time counterparts have cubic and quadratic forms.  
Indeed, the exact theoretical expressions given by 
Eqs.~(\ref{ctilinf})-(\ref{dtilinf}) reproduce the numerical outcomes, 
validating our approach as furnishing meaningful tests for O-U models. 
However, for $a_1 \ge 1$, rich pictures for the low order approximations of $\tilde{D}_3$ 
and $\tilde{D}_4$ arise,  which hinder the asymptotic estimation.

For the general case with $b_2\ge 0$, we computed the third-order $\tau$-expansions 
for $\tilde{D}_3(x,\tau)$ and $\tilde{D}_4(x,\tau)$. 
Each power of $\tau$ of order $j\le3$ has pre-factor denoted by $\tilde{D}_3^{(j)}$ 
and $\tilde{D}_4^{(j)}$ respectively. We find, for $\tilde{D}_3$:   
\begin{eqnarray} \nonumber  
\tilde{D}_3^{(0)} &=& 0 \,,\\ \nonumber  
\tilde{D}_3^{(1)} &=&  -b_0 \alpha x -b_2\alpha x^3 \,,\\ \nonumber  
\tilde{D}_3^{(2)} &=& \frac{1}{6}b_0 \alpha (9a_1-16b_2) x 
-\frac{1}{6} \alpha(a_1^2-13a_1b_2+16b_2^2) x^3 \,,\\  \label{D3til}
\tilde{D}_3^{(3)} &=& -\frac{1}{6}b_0 \alpha^2 (8a_1-13b_2) x   
+\frac{1}{12} \alpha(3a_1^3-32a_1^2b_2+74a_1b_2^2-52b_2^3) x^3   \,,
\end{eqnarray}
where $\alpha=a_1-2b_2$. At all orders, $\tilde{D}_3$ vanishes if $a_1=2b_2$.

\begin{figure}[h!]
\centering
\includegraphics*[bb=110 30 520 780, width=0.6\textwidth]{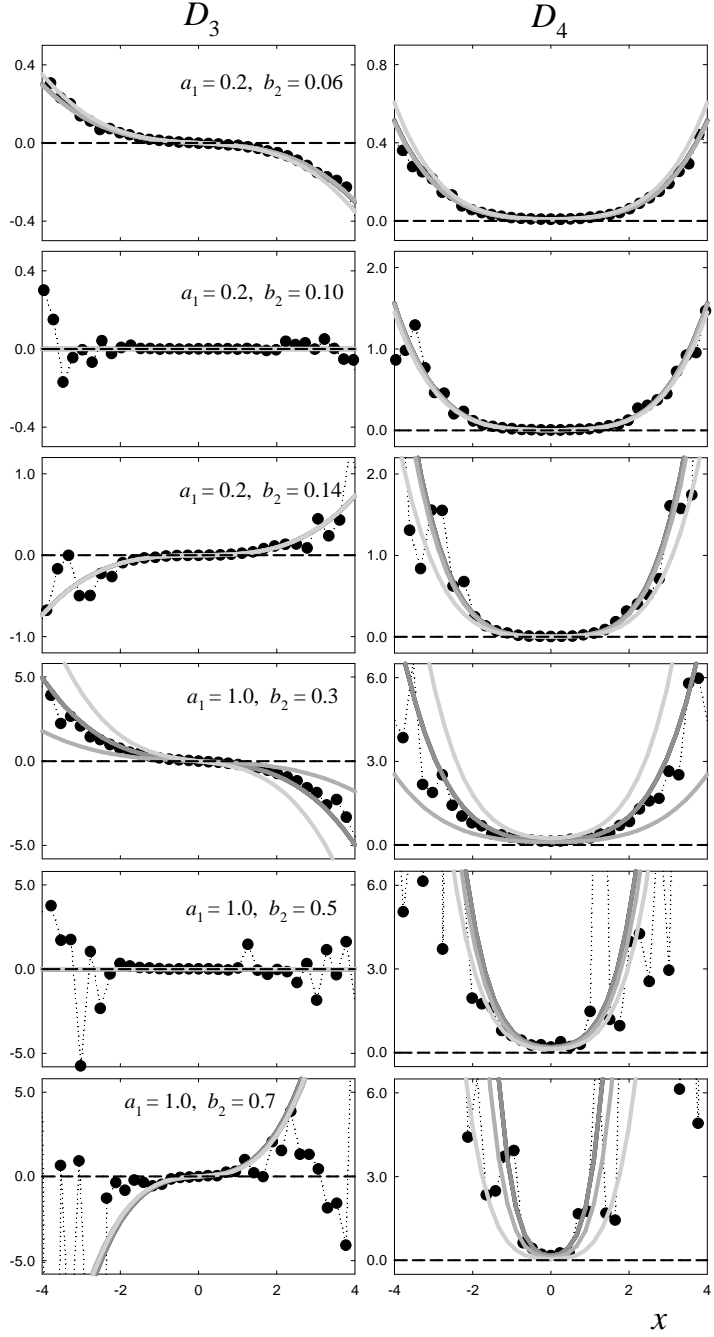}  
\caption{Third and fourth order coefficients for the general process with multiplicative noise. 
Symbols correspond  to the numerical computation for the same artificial series of 
Fig.~\ref{fig:b2D1D2}. 
Lines represent the coefficients given by Eqs.~(\ref{D3t})-(\ref{D4t}), using 
the theoretical $\tau$-expansions~(\ref{D3til})-(\ref{D4til}), 
at different orders of truncation up to third order. Colors as in previous figures.  
}
\label{fig:b2D3D4}
\end{figure}

For $\tilde{D}_4$, we have:
\begin{eqnarray} \nonumber 
\tilde{D}_4^{(0)} &=& 0 \,,\\ \nonumber  
\tilde{D}_4^{(1)} &=&  \frac{1}{2}b_0^2 +b_0b_2x^2+\frac{1}{2}b_2^2x^4 \,,\\ \nonumber  
\tilde{D}_4^{(2)} &=&  -\frac{1}{3}b_0^2(3a_1-7b_2) 
+\frac{1}{6}b_0(3a_1^2-30a_1b_2+52b_2^2)x^2 
 +\frac{1}{6}b_2(3a_1^2-24a_1b_2+38b_2^2)x^4 \,, \\ \nonumber
\tilde{D}_4^{(3)} &=& \frac{1}{6}b_0^2(7a_1^2-34a_1b_2+43b_2^2)  
 -\frac{1}{6}b_0(6a_1^3-65a_1^2b_2+206a_1b_2^2
-206b_2^3)x^2 \\      \label{D4til} 
&&+\frac{1}{24} (a_1^4-36a_1^3b_2+276a_1^2b_2^2-736a_1b_2^3  
 +652b_2^4)x^4   \,.
\end{eqnarray}

In Fig.~\ref{fig:b2D3D4}, we show the numerical computation of $\tilde{D}_3$ 
and $\tilde{D}_4$ for the same artificial series as in Fig.~\ref{fig:b2D1D2}. 
According to our theoretical results, $a_1=2b_2$ is a threshold between positive and 
negative slopes of $\tilde{D}_3$, as illustrated in Fig.~\ref{fig:b2D3D4}. 
For comparison, we also show up to the third order theoretical estimates 
$\sum_{j\ge 1} \tilde{D}_3^{(j)}$ and  $\sum_{j\ge 1} \tilde{D}_4^{(j)}$, 
according to Eqs.~(\ref{D3til})-(\ref{D4til}). 

Coefficientes $\tilde{D}_3$ and $\tilde{D}_4$ provide further tests of validity of the diffusive modeling. 
From Fig.~\ref{fig:b2D1D2}, third order estimates 
furnish suitable forecast of the numerical finite-$\tau$ measurements for small enough $a_1$.
In such cases, once obtained $a_1,\;b_0$ from Eqs.~(\ref{a1})-(\ref{b0}), these values can be used in 
theoretical equations for $\tilde{D}_3$ and $\tilde{D}_4$, to check if the corresponding 
non-null coefficients can be attributed to finite-$\tau$ effects.

\section{Summary and final comments}

For an important class of  diffusion models with additive-multiplicative noise, 
we have derived exact formulas that connect the empirical  
discrete-time estimates with the actual 
values of the parameters of drift and diffusion coefficients. 
Additionally, we also provided theoretical expressions for higher-order coefficients which serve as a 
further probe for the validity of this class of diffusive models.

Our results allow to access the generating stochastic process. 
A possible procedure to identify it and its parameters can be summarized as follows. 
When numerical computation of the coefficients from a real timeseries 
yields linear and quadratic forms for $D_1$ and $D_2$ (which is a frequent outcome), 
 the values of $\{\tilde{a}_1,\tilde{b}_0,\tilde{b}_2\}$ can be obtained from curve fitting.  
The present model (with or without multiplicative component) would be  adequate when  
i) $\tilde{a}_1<1$ and ii) ($\tilde{a}_1\simeq\tilde{b}_0+\tilde{b}_2$).
For  $\tilde{a}_1<0.5$ the second order correction would be enough to recover $a_1$, 
otherwise larger order corrections 
should be considered. Once $\tilde{a}_1$ and $\tilde{b}_0$ are known, 
Eqs.~(\ref{a1})  and (\ref{b0}), allow to obtain, 
exactly, the original parameters $a_1,b_0$, hence also $b_2=a_1-b_0$. 
Values such that  $b_2<<a_1,b_0$ (hence, $a_1\simeq b_0$) point to a simple O-U process, 
otherwise a multiplicative term may be also present. 
In both cases, a further check consists in the analysis of higher order coefficients, e.g., 
to see whether a non-null $D_4$ can be attributed to finite-time corrections.

By analyzing the O-U process, we also found that a low sampling rate would 
significantly affect the diffusion coefficient estimate, by adding an extra 
quadratic term.
Thus, the detection of a quadratic $\tilde{D}_2$ does not imply the existence of 
multiplicative components in the actual process. 
Moreover, estimations of $D_3$ and $D_4$ from low-order approximations would lead to 
 results inconsistent with the empirical outcomes.

The obtained formulas also allow to quantify the errors induced by a finite 
sampling rate $\tau$ in the numerically estimated coefficients.
The analytical results indicate that, in order to grasp the  
true values of the parameters from the knowledge of the observed ones,  
the required correction depends strongly on the (hidden) inverse time $a_1$. 
Our work shows that one should be careful when applying low-order finite-$\tau$ 
corrections for diffusion models.
Furthermore, as shown in Fig.~\ref{fig:orders}, our results provide a criterion, 
from  the knowledge of $\tilde{a}_1$, 
to determine the required order $n$,  or equivalently, up to which value of $\tau$ the respective 
approximation is reliable.  

\section*{Acknoledgements:}
We acknowledge Brazilian agencies Faperj and CNPq for partial financial support.

\section*{Appendix}

Rewriting Eq.~(\ref{ABtilde}) according to the notation introduced in Eq.~(\ref{MSI}), one has
\begin{equation}  \nonumber
\tau\tilde{D}_1=\langle \Delta X\rangle = \sum_{\alpha_k} c_{\alpha_k}(D_1,D_2)\langle I_{\alpha_k}\rangle. 
\end{equation}
Only multiple stochastic integrals $I_{\alpha_k}$ such that $\alpha_k=(0,...,0)_k$ have non-null 
average, being $\langle I_{(0,...,0)_k}\rangle =\tau^k/k!$. 
From the iterated application of It\^o formula to Eq.~(\ref{intXform}), 
$c_{(0,...,0)_k}(D_1,D_2)=(L^0)^{k-1} D_1$.
These are general results independent of the particular form of $D_1$ and $D_2$.
Noticing that, from Eq.~(\ref{L0L1}), $L^0=\partial_t+D_1\partial_x+D_2\partial_{xx}$, 
and that $D_1$ is time-independent and linear in $x$,  
then, $c_{\alpha_k}=D_1(D_1')^{k-1}=(-a_1)^{k} x$. Finally,   
\begin{equation} \nonumber
\tilde{D}_1=\langle\Delta X\rangle/\tau =\sum_{k\ge 1} \frac{1}{k!}(-a_1)^k\tau^{k-1}x\,,
\end{equation}
which is of the same functional form of the true $D_1$ and can be identified 
with $-\tilde{a}_1x$, so that  
\begin{equation} \nonumber
\tilde{a}_1=  -\sum_{k\ge 1} \frac{1}{k!}(-a_1)^k\tau^{k-1}\,,
\end{equation}
which gives Eq.~(\ref{a1n}).

For the second conditional moment, one has
\begin{equation}  \label{Ap_D2til}
2\tau \tilde{D}_2 = \langle (\Delta X)^2 \rangle = \sum_{\alpha_n,\beta_m} c_{\alpha_n}c_{\beta_m}
\langle I_{\alpha_n}I_{\beta_m}\rangle. 
\end{equation}
From the definition of $c_{\alpha_k}$ in Eq.~(\ref{MSI}),  
if $b_2=0$ (then $D_2$ is constant),  
only two classes of terms in Eq.~(\ref{Ap_D2til}) are non-null, those with:\\
 i) $\alpha_n=(0,...,0)_n$ and $\beta_m=(0,...,0)_m$ and \\
 ii) $\alpha_n=(1,0,...,0)_n$ and $\beta_m=(1,0,...,0)_m$. 
In those cases, the products $c_{\alpha_n}c_{\beta_m}$ take the values: \\
i) $(D_1)^2(D_1')^{k-1}=(-a_1)^{k+1}x^2$ (with $k=m+n-1$), \\
ii) $2D_2(D_1')^{k}=2b_0 (-a_1)^{k}$ (with $k=m+n-2$).
\\ \noindent
In order to evaluate the averages of products of multiple stochastic 
integrals, it is useful to recall that   
$\langle I_{(0,...,0)_n}I_{(0,...,0)_m}\rangle=\frac{\tau^{n+m}}{n!m!}$ and that
$\langle I_{(1,...,0)_n}I_{(1,...,0)_m}\rangle=
\frac{\tau^{n+m-1}}{(n+m-1)(n-1)!(m-1)!}$~\cite{book}.   
Then, summing over all the pairs $(n,m)$ contributing to the order 
$\tau^{k+1}$, one obtains: \\  
i) $\sum \langle I_{\alpha_n}I_{\beta_m}\rangle/\tau^{k+1}= 
\frac{1}{(k+1)!} \sum_{n=1}^k \binom{k+1}{n}
=2\frac{2^k-1}{(k+1)!}$, \\ 
ii) $\sum\langle I_{\alpha_n}I_{\beta_m}\rangle/\tau^{k+1}=
\frac{1}{(k+1)!} \sum_{j=0}^k \binom{k}{j}
=\frac{2^k}{(k+1)!}$. 
\\ \noindent
Finally, from Eq.~(\ref{Ap_D2til}), we arrive at 
\begin{equation}
\tilde{D}_2=\langle(\Delta X)^2\rangle/(2\tau) 
=\frac{1}{2}\sum_{k\ge 0}\biggl(  
2\frac{2^k-1}{(k+1)!}(-a_1)^{k+1}x^2 +
\frac{2^k}{(k+1)!}2b_0(-a_1)^k   
\biggr)\tau^k \,, 
\end{equation}
which can be cast in the form $\tilde{b}_2x^2+\tilde{b}_0$, allowing to 
identify $\tilde{b}_2$ and $\tilde{b}_0$ with functions of the true parameters, as
\begin{eqnarray} \nonumber
\tilde{b}_0 &=&  \frac{1}{2} \sum_{k\ge 0} \frac{2^k}{(k+1)!}2b_0(-a_1)^k \tau^k\,, 
\\ \label{bs} 
\tilde{b}_2 &=&  \sum_{k\ge 0}\frac{2^k-1}{(k+1)!}(-a_1)^{k+1} \tau^k \,. 
\end{eqnarray}

For the general case $b_2\ge 0$, a similar but tricky derivation  leads to 
Eqs.~(\ref{b0n})-(\ref{b2n}) that generalize the expressions (\ref{bs}).

Proceeding with the third order, products of three multiple integrals appear. For $b_2=0$, 
there are two types of products  $I_{\alpha_n}I_{\beta_m}I_{\gamma_l}$ 
contributing to $\langle(\Delta X)^3\rangle$: \\
i) $\alpha_n=(0,...,0)_n$, $\beta_m=(0,...,0)_m$, $\gamma_l=(0,...,0)_l$ and \\
ii) $\alpha_n=(0,...,0)_n$, $\beta_m=(1,0,...,0)_m$, $\gamma_l=(1,0,...,0)_l$, with 
pre-factors proportional to $(D_1)^3(D_1')^{k-2}=(-a_1)^{k+1}x^3$ and 
to $2D_2D_1(D_1')^{k-1}=2b_0(-a_1)^{k}x$, respectively. 
Evaluating the averages of products of three multiple stochastic 
integrals and summing over all the triplets $(n,m,l)$ contributing to the same 
order in  $\tau$, as done for the second order coefficient, 
one gets $\tilde{D}_3=-\tilde{c}_1(\tau)+ \tilde{c}_3(\tau)x^3$, as in Eq.~(\ref{D3t}).

Analogously, at fourth order, considering the relevant products of four 
multiple integrals, for $b_2=0$, 
three types of contributions appear, 
yielding $\tilde{D}_4(x,\tau) = \tilde{d}_0(\tau)+\tilde{d}_2(\tau)x^2 + \tilde{d}_4(\tau)x^4$, 
as in Eq.~(\ref{D4t}). 

Let us recall that the averages of products of $n$ multiple stochastic integrals appearing in 
the $nth$-order term of the  
coefficients can be expressed, in general, as multinomial terms, whose summation over all the products 
has the form $\mu_11^k+\mu_22^k+\mu_3 3^k +....\mu_n n^k$ (with rational $\mu_i$) for the order $k$ in $\tau$.
Moreover, products of multiple stochastic integrals can be readily simplified, by means of useful relations 
between multiple It\^o integrals~\cite{book}. 
Then, although at the cost of increasing the number of indices, the number of factors can be reduced.

\end{document}